%% file: ASILOMAR_Conference/ASILOMAR_Paper.tex
\documentclass[10pt,conference,romanappendices]{IEEEtran}
\IEEEoverridecommandlockouts
\usepackage{algorithm,algorithmic,amsmath,amssymb,amsthm,bbm,cite,color,comment,enumitem,graphicx,microtype,subcaption,setspace,url}
\usepackage{mathtools,xparse}
\usepackage[USenglish]{babel}
\usepackage{hyperref}
\usepackage[utf8]{inputenc}
\usepackage[T1]{fontenc}
\usepackage{float}
\usepackage[autostyle]{csquotes}
\usepackage[labelsep=period,font=small]{caption}
\usepackage[nolist]{acronym}
\usepackage[normalem]{ulem}

\usepackage{tikz,pgfplots}
\usetikzlibrary{arrows, arrows.meta, backgrounds, calc, intersections, decorations.markings, decorations.pathreplacing, positioning, shapes}
\pgfplotsset{compat=1.17}
\usepackage{caption}
\usepackage{acronym}

\acrodef{los}[LoS]{\emph{Line of sight}}
\acrodef{nlos}[NLoS]{\emph{Non-line of sight}}

\input{./Definitions.tex}
\input{./Acronyms.tex}

\usepackage{titlesec}
\let\subparagraph\relax
\titlespacing{\section}{0pt}{8pt plus 2pt minus 1pt}{6pt plus 1pt minus 1pt} 
\titlespacing{\subsection}{0pt}{7pt plus 2pt minus 1pt}{5pt plus 1pt minus 1pt} 
\title{Wavenumber-Division Multiplexing in Holographic MIMO with NLoS Channels}
\author{
\IEEEauthorblockN{Ashutosh Prajapati,$^{1}$ Prathapasinghe Dharmawansa,$^{1}$ Marco Di Renzo,$^{2,3}$ and Italo Atzeni$^{1}$}
\IEEEauthorblockA{$^{1}$ Centre for Wireless Communications, University of Oulu, Finland \\
$^{2}$ Université Paris-Saclay, CNRS, CentraleSupélec, Laboratoire des Signaux et Systèmes, France \\
$^{3}$ King's College London, Centre for Telecommunications Research, Department of Engineering, United Kingdom \\
Emails: \{ashutosh.prajapati, prathapasinghe.kaluwadevage, italo.atzeni\}@oulu.fi, marco.di\_renzo@kcl.ac.uk
\thanks{The work of A.~Prajapati, P.~Dharmawansa, and I.~Atzeni was supported by the Research Council of Finland (336449 Profi6, 348396 HIGH-6G, and 369116 6G~Flagship). The work of M. Di Renzo was supported by the France-Nokia Chair of Excellence in ICT, by the European Union through the Horizon Europe projects COVER (101086228), UNITE (101129618), INSTINCT (101139161), and TWIN6G (101182794), by the Agence Nationale de la Recherche (ANR) through the France 2030 project ANR-PEPR Networks of the Future (NF-Founds 22-PEFT-0010), by the CHIST-ERA project PASSIONATE (CHIST-ERA-22-WAI-04 and ANR-23-CHR4-0003-01), and by the Engineering and Physical Sciences Research Council (EPSRC), part of UK Research and Innovation, and the UK Department of Science, Innovation and Technology through the CHEDDAR Telecom Hub (EP/X040518/1 and EP/Y037421/1) and through the HASC Telecom Hub (EP/X040569/1).}}}

\begin{document}

\maketitle

\begin{abstract}
Wavenumber-division multiplexing (WDM) was introduced as a counterpart of orthogonal frequency-division multiplexing in the spatial-frequency domain for line-of-sight holographic multiple-input multiple-output (MIMO) systems. In this paper, we extend WDM to holographic MIMO channels with non-line-of-sight (NLoS) propagation. We show that applying WDM to the NLoS channel yields the corresponding angular-domain representation, which we characterize through the power spectral factor and power spectral density. We further obtain a closed-form characterization for the case of isotropic scattering, recovering Jakes' isotropic model. The analysis is complemented by numerical results evaluating the degrees of freedom and ergodic capacity under both isotropic and non-isotropic scattering.
\end{abstract}


\section{Introduction} \label{sec:Intro}

Wireless data demand is growing at an unprecedented pace, pushing communications systems toward higher frequencies where extremely dense deployments of physically small antennas become feasible \cite{Atz25}. This trend has sparked strong interest in holographic \ac{MIMO} technologies as a means to support future high-capacity networks and intelligent applications. Unlike conventional massive \ac{MIMO}, where antenna elements are typically spaced about half a wavelength apart to limit spatial correlation \cite{wei2024electromagnetic}, holographic \ac{MIMO} envisions continuous apertures (i.e., surfaces or lines) or densely packed antenna arrays, which become electrically large. In this regime, near-field propagation becomes dominant, in contrast to classical massive \ac{MIMO} that generally operates under far-field conditions \cite{sharma2021interference}. Holographic \ac{MIMO} is often realized using programmable metamaterials \cite{huang2020holographic}, enabling flexible hardware architectures and multifunctional operation across time, frequency, and space.

Continuous-aperture \ac{MIMO} models arise either from the spatial sampling of the transmit and receive apertures, or from expanding the transmit current and received field using suitable basis functions. The latter approach yields the \ac{WDM} framework, introduced in \cite{sanguinetti2022wavenumber} and inspired by \ac{OFDM}: while \ac{OFDM} operates in the frequency domain, \ac{WDM} enables multiplexing in the wavenumber (spatial-frequency) domain through an orthogonal decomposition of the continuous current and field. Prior \ac{WDM} studies build on a holographic line model under purely \ac{LoS} propagation \cite{sanguinetti2022wavenumber, d2022performance}, whereas \ac{NLoS} propagation has been explored only for spatial-sampling-based frameworks \cite{pizzo2022spatial, pizzo2020holographic}. Although millimeter-wave and sub-terahertz systems are often \ac{LoS}-dominated, small-scale fading remains significant, so \ac{NLoS} components cannot be entirely neglected.

In this paper, we extend the \ac{WDM} framework using the \ac{NLoS} holographic line model introduced in \cite{praj25journal} and show that applying \ac{WDM} to the \ac{NLoS} channel yields the corresponding angular-domain representation. By analyzing the spatial autocorrelation of the \ac{WDM}-based \ac{NLoS} channel, we establish a relation between the \ac{PSF} and \ac{PSD}, enabling an explicit characterization of the resulting angular-domain channel. We further derive closed-form expressions of the \ac{ACF} and \ac{PSD} under isotropic scattering, which allows us to retrieve the classical Jakes' isotropic model. The results demonstrate that \ac{WDM} serves not only as a multiplexing technique but also as a powerful analytical tool for investigating holographic \ac{MIMO} channels. Finally, we evaluate the \ac{DoF} and ergodic capacity of the considered \ac{WDM}-based \ac{NLoS} channel and show that both degrade significantly under non-isotropic scattering compared with the isotropic case.

\section{System Model} \label{sec:Sys}

\begin{figure}[t]
    \centering
\input{ASILOMAR_Conference/ASILOMAR_Results/files_tex/WDM_ASILOMAR}
    \caption{A schematic of the considered \ac{WDM}-based \ac{NLoS} holographic \ac{MIMO} system model.}
    \label{fig:WDM System Model}
\end{figure}
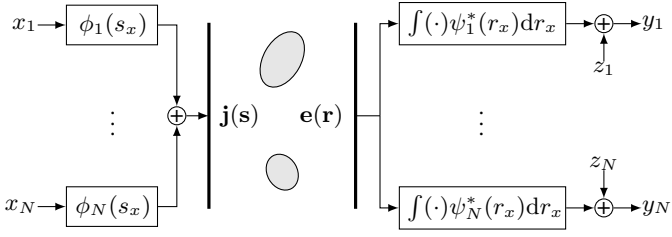

Consider a \ac{WDM}-based holographic \ac{MIMO} system over an \ac{NLoS} channel as depicted in Fig.~\ref{fig:WDM System Model}, where a line source spanning the linear region $\setL_{\textrm{s}} \subset \Real^{2}$ with length $L_{\textrm{s}}$ transmits data to a line receiver spanning the linear region $\setL_{\textrm{r}} \subset \Real^{2}$ with length $L_{\textrm{r}}$. The two lines are parallel and oriented along the $x$-axis, with their centers aligned along the $z$-axis and separated by a distance $d$. Let $\s = [s_{x}, s_{z}]^{\tran} \in \setL_{\textrm{s}}$ and $\r = [r_{x}, r_{z}]^{\tran} \in \setL_{\textrm{r}}$ denote arbitrary points within the source and receiver regions, respectively. Throughout the paper, we assume that the communication takes place via scalar waves (as in, e.g., \cite{pizzo2022spatial,pizzo2020holographic}), which simplifies the analysis by allowing the use of the scalar Green's function. The detailed \ac{EM}-based channel model for holographic lines is presented in \cite{praj25journal}.
Let $\kappab=[\kappa_{x},\gamma(\kappa_{x})]^{\tran} \in \Real^2$ and $\k=[k_{x}, \gamma(k_{x})]^{\tran} \in \Real^2$ be the wave vectors corresponding to the transmit and receive propagation directions $\hat{\kappab} =\frac{\kappab}{\|\kappab\|}$ and $\hat{\k} =\frac{\k}{\|\k\|}$, respectively. The corresponding wavenumbers are defined as $\kappa = k = \frac{2\pi}{\lambda}$, where $\lambda$ is the wavelength. Following \cite{praj25journal}, the \ac{EM}-based \ac{NLoS} channel impulse response is given by  
\begin{align}
\label{eqn: LoS Plane Wave}
h(\r,\s)=\frac{1}{2\pi}\int_{\Real^2}a_{\textrm{r}}(\k,\r)H_{\textrm{a}}(k_{x}, \kappa_{x})a_{\textrm{s}}(\kappab,\s)\diff k_{x}\diff\kappa_{x},
\end{align}
where $H_{\textrm{a}}(k_{x}, \kappa_{x})$ represents the corresponding angular-domain channel impulse response (defined later) and
\begin{subequations}
\label{eqn:plane waves source and receiver}
\begin{align}
a_{\textrm{s}}(\kappab,\s) & =e^{-j\kappab^{\tran}\s}=e^{-j(\kappa_{x}s_{x}+\gamma(\kappa_{x})s_{z})}, \\
a_{\textrm{r}}(\k,\r) & =e^{j\k^{\tran}\r}=e^{j(k_{x}r_{x}+\gamma(k_{x})r_{z})}
\end{align}
\end{subequations}
are the transmit and receive plane waves, respectively.

In the presence of scatterers between the holographic lines, the one-to-one correspondence between the plane waves at the source and receiver no longer holds. Let $A(k_{x}, \kappa_{x})$ be a non-negative function characterizing the scattering environment, referred to as the \ac{PSF}, and let $W(k_{x}, \kappa_{x})$ denote a spatially stationary complex white Gaussian noise random field. In particular, the \ac{PSF} describes the coupling strength between the source and receiver. Under this model, the angular-domain channel is given by \cite{pizzo2022spatial}
\begin{align}
\label{eqn: Angular LoS}
H_{\textrm{a}}(k_{x}, \kappa_{x}) = \frac{A(k_{x}, \kappa_{x}) W(k_{x}, \kappa_{x})}{\sqrt{\gamma(k_{x})\gamma(\kappa_{x})}}, \ (k_{x}, \kappa_{x}) \in \setD^{2},
\end{align}
with $\setD=\{\kappa_{x}\in \Real: -\kappa\leq\kappa_{x}\leq \kappa\}$ and
\begin{align}
\label{eqn:gamma}
\gamma(\kappa_{x})=\sqrt{\kappa^{2}-\kappa_{x}^{2}}, \ \kappa_{x}\in \setD
\end{align}
for the traveling waves. Hence, plugging \eqref{eqn: Angular LoS} into \eqref{eqn: LoS Plane Wave}, the \ac{NLoS} channel impulse response can be written as
\begin{align}
\label{eqn:NLoSSpatial}
h(\r,\s)= \, & \nonumber\frac{1}{2\pi}\int_{\setD^{2}}a_{\textrm{r}}(\k,\r)\frac{A(k_{x}, \kappa_{x})W(k_{x}, \kappa_{x})}{\sqrt{\gamma(k_{x})\gamma(\kappa_{x})}}\\
& \times a_{\textrm{s}}(\kappab,\s)\diff k_{x}\diff\kappa_{x},
\end{align}
where $\setD^{2}$ specifies the support of the channel in the continuous wavenumber domain. Since $h(\r,\s)$ is bandlimited within $\setD^{2}$, sampling this support as in \cite{pizzo2020holographic} yields its discretized counterpart formed by the sets
\begin{subequations}
\label{eqn: transmit set and receive set}
\begin{align}
\setE_{\textrm{s}}& =\left\{p_x \in \mathbb{Z}:-\kappa \leq \frac{2\pi}{L_{\textrm{s}}}p_x\leq \kappa\right\},\\
\setE_{\textrm{r}}& =\left\{q_x \in \mathbb{Z}:-k \leq \frac{2\pi}{L_{\textrm{r}}}q_x\leq k \right\}
\end{align}
\end{subequations}
at the source and receiver, respectively. Now, define $ n_{\textrm{s}} = \card(\setE_{\textrm{s}}) =\big\lfloor\frac{2 L_{\textrm{s}}}{\lambda}\big\rfloor$ and $n_{\textrm{r}} = \card(\setE_{\textrm{r}}) =\big\lfloor\frac{2 L_{\textrm{r}}}{\lambda}\big\rfloor$ \cite{pizzo2020degrees}. Following the Fourier plane-wave series expansion of a spatially stationary complex Gaussian random process from \cite{pizzo2020holographic}, we have
\begin{align}
\label{eqn: Discrete NLoS}
h(\r,\s)\approx\sum_{q_x\in\setE_{\textrm{r}}}\sum_{p_x \in\setE_{\textrm{s}}} H_{\textrm{a}}({q_x, p_x})a_{\textrm{r}}(q_x, \r)a_{\textrm{s}}(p_x, \s),
\end{align}
where
\begin{subequations}
\label{eqn: Discretize Plane wave source and receiver}
\begin{align}
a_{\textrm{s}}(p_x,\s)&=e^{-j\left(\frac{2\pi}{L_{\textrm{s}}}p_x s_{x}+\gamma(p_x)s_{z}\right)},\\
a_{\textrm{r}}(q_x,\r)&=e^{j\left(\frac{2\pi}{L_{\textrm{r}}}q_xr_{x}+\gamma(q_x)r_{z}\right)}
\end{align}
\end{subequations}
are the discretized counterparts of \eqref{eqn:plane waves source and receiver}, with $\gamma(p_x)=\sqrt{\kappa^{2}-\big(\frac{2\pi p_x}{L_{\textrm{s}}}\big)^2}$ and $\gamma(q_x)=\sqrt{k^{2}-\big(\frac{2\pi q_x}{L_{\textrm{r}}}\big)^2}$, and
\begin{align}
    \label{eqn: Discrete Angular}
    H_{\textrm{a}}(q_x, p_x)\sim \setN_{\mathbb{C}}(0,\sigma^2(q_x, p_x))
\end{align}
acts as a coupling coefficient characterizing the interaction between the transmit and receive plane waves.

\section{WDM with EM-Based NLoS Channel}
\label{sec: WDM for NLoS}

In this section, we first study the \ac{WDM}-based \ac{NLoS} channel in Section~\ref{sec: WDM-applied NLoS Channel}, then derive its spatial autocorrelation properties in Section~\ref{sec: Autocorrelation of Channel H_{nm}}, and finally characterize the resulting angular-domain channel in Section~\ref{sec: Calculation of Channel Coefficients for WDM and holographic channel}. For simplicity, we set $s_z=0$ and $r_z=d$; consequently, the source and receiver span the linear regions $\setL_{\textrm{s}}=\{(s_x, 0): |s_x|\leq \frac{L_{\textrm{s}}}{2}\} $ and $\setL_{\textrm{r}}= \{(r_x, d): |r_x|\leq \frac{L_{\textrm{r}}}{2}\}$, respectively.

\subsection{WDM-Based NLoS Channel}
\label{sec: WDM-applied NLoS Channel}

Let $\{x_m\}_{m=1}^{N}$ denote the transmit data symbols, with $N \leq \mathrm{min}(n_{\textrm{s}}, n_{\textrm{r}})$. The electric current at the source (measured in amperes) is constructed as $i(s_x) = \sum_{m=1}^N x_m \phi_m(s_x)$, where $\big\{\phi_m(s_x)\big\}_{m=1}^{N}$ represent the transmit Fourier basis, with $m$-th basis function
\begin{align}
\label{eqn: basis function at source}
   \phi_m(s_x)= 
\begin{cases}
    \frac{1}{\sqrt{L_{\textrm{s}}}}e^{j\frac{2\pi}{L_{\textrm{s}}}m s_x},&  |s_x|\leq \frac{L_{\textrm{s}}}{2}\\
    0,              & \textrm{otherwise},
\end{cases}
\end{align}
and $x_m=\int_{-\frac{L_{\textrm{s}}}{2}}^{\frac{L_{\textrm{s}}}{2}} i(s_x)\phi_m^*(s_x) ds_x$. The current density at the source is defined as
\begin{align}
\label{eqn: current distribution}
   \j(\s)=i(s_x)\delta(s_z)\hat{\x},
\end{align}
where $\hat{\x} = [1, 0, 0]^\tran$ is the unit vector along the $x$-axis.
The electric field at the receive point $\r$ given the current density at the source in \eqref{eqn: current distribution} is expressed as \cite{pizzo2020holographic}
\begin{align}
\begin{split}
   \e(\r) & =\int_{-\infty}^{\infty} h(\r, \s)\,\j(\s)\diff\s \\
    & =\int_{-\frac{L_{\textrm{s}}}{2}}^{\frac{L_{\textrm{s}}}{2}} h(\r, \s) \sum_{m=1}^N\hspace{-1mm} \xi_m\phi_m(s_x) \hat{\x} \diff s_x \\
    & = \big[e_{x}(r_{x}), 0, 0 \big]^{\tran},
\end{split} 
\label{eqn: convolution spatially variant} 
\end{align}
with $h(\r, \s)$ defined in \eqref{eqn:NLoSSpatial}. Then, $\e(\r)$ is projected onto the inner-product space spanned by
\begin{align}
    \label{eqn: receiver vector space}
    \psib_n(\r)= \psi_n(r_x)\delta(r_z-d)\hat{\x},
\end{align}
where $\big\{\psi_n(r_x)\big\}_{n=1}^{N}$ represent the receive Fourier basis, with $n$-th basis function
\begin{align}
\label{eqn: basis function at receiver}
\psi_n(r_x)=
\begin{cases}
    \frac{1}{\sqrt{L_{\textrm{r}}}}e^{j\frac{2\pi}{L_{\textrm{r}}}n r_x},&  |r_x|\leq \frac{L_{\textrm{r}}}{2}\\
    0,              & \textrm{otherwise}.
\end{cases}
\end{align}
Hence, the $n$-th spatial sample of the received signal is given~by
\begin{align}
\label{eqn: spatial sample at receiver}
  y_n=\int_{-\frac{L_{\textrm{r}}}{2}}^{\frac{L_{\textrm{r}}}{2}} \psi_n^*(r_x) e_{x}(r_{x}) \diff r_x+ z_n,
\end{align}
where $z_n \in \setN_{\Compl}(0, \chi^2)$ is the additive white Gaussian noise term with variance $\chi^2$. Finally, plugging \eqref{eqn: convolution spatially variant} into \eqref{eqn: spatial sample at receiver} yields
 \begin{align}
\label{eqn: input-output relationship}
  y_n=\sum_{m=1}^{N} H_{n, m} x_m+ z_n,
\end{align}
where
\begin{align}
\label{eqn: Channel H_nm mod}
 H_{n, m}=\int_{-\frac{L_{\textrm{r}}}{2}}^{\frac{L_{\textrm{r}}}{2}}\int_{-\frac{L_{\textrm{s}}}{2}}^{\frac{L_{\textrm{s}}}{2}}\psi_n^*(r_x) h(\r, \s) \phi_m(s_x) \diff s_x\diff r_x
\end{align}
is the \ac{WDM}-based \ac{NLoS} channel corresponding to the $m$-th transmit and $n$-th receive Fourier basis functions.

Combining \eqref{eqn: Channel H_nm mod} and \eqref{eqn:NLoSSpatial} leads to
\begin{align}
\label{eqn: Channel H_nm}
 H_{n, m}=\, &  \frac{1}{2\pi}\int_{-\frac{L_{\textrm{r}}}{2}}^{\frac{L_{\textrm{r}}}{2}}\int_{-\frac{L_{\textrm{s}}}{2}}^{\frac{L_{\textrm{s}}}{2}}\int_{\Real^2}\psi_n^*(r_x)a_{\textrm{r}}(\k,\r) \frac{A(k_x, \kappa_x)}{\sqrt{\gamma(k_x)}}\nonumber  \\ &\times \frac{W(k_x, \kappa_x)}{\sqrt{\gamma(\kappa_x)}} a_{\textrm{s}}(\kappab,\s) \phi_m(s_x) \diff k_x\diff\kappa_x \diff s_x\diff r_x,
\end{align}
which is used in Section~\ref{sec: Autocorrelation of Channel H_{nm}} to derive the spatial autocorrelation of $H_{n,m}$. Now, we substitute the Fourier series expansion of $h(\r, \s)$, given in \eqref{eqn: Discrete NLoS} with angular-domain channel impulse response in \eqref{eqn: Discrete Angular}, into \eqref{eqn: Channel H_nm mod}. Utilizing \eqref{eqn: Discretize Plane wave source and receiver} along with the fact that $H_{\textrm{a}}(q_x, p_x)$ and $e^{j\frac{2\pi}{L_{\textrm{r}}}\gamma(q_x)r_z} H_{\textrm{a}}(q_x, p_x)e^{-j\frac{2\pi}{L_{\textrm{s}}}\gamma(p_x)s_z}$ are statistically equivalent, $H_{n, m}$ can be expressed with a slight abuse of notation as
\begin{align}
\begin{split}
 H_{n, m}=\, & \frac{1}{\sqrt{L_{\textrm{r}} L_{\textrm{s}}}}\int_{-\frac{L_{\textrm{r}}}{2}}^{\frac{L_{\textrm{r}}}{2}}\int_{-\frac{L_{\textrm{s}}}{2}}^{\frac{L_{\textrm{s}}}{2}}\sum_{q_x\in\setE_{\textrm{r}}}\sum_{p_x \in\setE_{\textrm{s}}} H_{\textrm{a}}({q_x, p_x})\\
 &\times e^{j\frac{2\pi}{L_{\textrm{r}}}(q_x-n) r_x}e^{-j\frac{2\pi}{L_{\textrm{s}}} (p_x-m) s_x} \diff s_x\diff r_x \\
 = \, & \sqrt{L_{\textrm{r}} L_{\textrm{s}}}\sum_{q_x\in\setE_{\textrm{r}}}\sum_{p_x \in\setE_{\textrm{s}}}H_{\textrm{a}}({q_x, p_x})\mathrm{sinc}(q_x-n) \\
 &\times \mathrm{sinc}(p_x-m) \\
 = \, & \sqrt{L_{\textrm{r}} L_{\textrm{s}}}H_{\textrm{a}}({n, m}), 
\end{split}
\label{eqn: Channel H_nm discrete without sinc}
\end{align}
with $n\in\setE_{\textrm{r}}$ and $m\in\setE_{\textrm{s}}$. The characterization of $H_{\textrm{a}}({n, m})$ is presented in Section~\ref{sec: Calculation of Channel Coefficients for WDM and holographic channel}. From \eqref{eqn: Channel H_nm discrete without sinc}, we observe that the channel resulting from applying \ac{WDM} corresponds to the angular-domain channel scaled by a gain that depends on the lengths of the holographic lines.

\subsection{Spatial Autocorrelation of \texorpdfstring{$H_{n, m}$}{}}
\label{sec: Autocorrelation of Channel H_{nm}}

In this section, we first derive the spatial autocorrelation of the \ac{WDM}-based \ac{NLoS} channel and then establish the relation between its \ac{PSD} and \ac{PSF}. The spatial autocorrelation of the \ac{WDM}-based \ac{NLoS} channel is given by
\begin{align}
    \label{eqn: Autocorrelation}
    R_{mn, qp}=\mathbb{E}\big[ H_{n, m} (H_{p, q})^*\big].
\end{align}
Since $W(k_x, \kappa_x)$ is a spatially stationary complex white Gaussian noise random field, we have $\mathbb{E}\big[W(k_x, \kappa_x)W^*(k'_x, \kappa'_x)\big]=\delta(k_x-k'_x)\delta(\kappa_x-\kappa'_x)$. Moreover, since $W(k_x, \kappa_x)$ and $e^{j\frac{2\pi}{L_{\textrm{r}}}\gamma(k_x)r_z} W(k_x, \kappa_x)e^{-j\frac{2\pi}{L_{\textrm{s}}}\gamma(\kappa_x)s_z}$ are statistically equivalent, plugging \eqref{eqn: Channel H_nm} into \eqref{eqn: Autocorrelation} leads to
\begin{align}
\label{eqn: Autocorrelation2}
 R_{mn, qp}=\,& \frac{1}{(2\pi)^2} \int_{-\frac{L_{\textrm{r}}}{2}}^{\frac{L_{\textrm{r}}}{2}}\int_{-\frac{L_{\textrm{s}}}{2}}^{\frac{L_{\textrm{s}}}{2}}\int_{-\frac{L_{\textrm{r}}}{2}}^{\frac{L_{\textrm{r}}}{2}}\int_{-\frac{L_{\textrm{s}}}{2}}^{\frac{L_{\textrm{s}}}{2}}\int_{\Real^2} \psi_n^*(r_x)\nonumber\\ &\times e^{jk_x(r_x-r'_x)}
 \psi_p(r'_x) \frac{A^2(k_x, \kappa_x)}{\gamma(k_x)\gamma(\kappa_x)}e^{-j\kappa_x (s_x-s'_x)}\nonumber \\ &\times \phi_m(s_x) 
  \phi^*_q(s'_x)\diff k_x\diff\kappa_x   \diff s_x\diff r_x \diff s'_x\diff r'_x.
\end{align}

Considering separable scattering at the source and receiver \cite{pizzo2022spatial}, we have $A^2(k_x, \kappa_x)=A_\textrm{r}^2(k_x)A_{\textrm{s}}^2(\kappa_x)$ and \eqref{eqn: Autocorrelation2} becomes
\begin{align}
\label{eqn: Autocorrelation Matrix elements}
[\R]_{mn, qp}=[\R_{\textrm{s}}]_{m, q} [\R_{\textrm{r}}]_{n, p},
\end{align}
where $\R_{\mathrm{s}} \in \mathbb{C}^{n_{\mathrm{s}} \times n_{\mathrm{s}}}$ and $\R_{\mathrm{r}} \in \mathbb{C}^{n_{\mathrm{r}} \times n_{\mathrm{r}}}$ denote the spatial autocorrelation matrices at the source and receiver, respectively, with
\begin{subequations}
\begin{align}
\label{eqn: Autocorrelation Matrix Source}
\big[\R_{\textrm{s}}\big]_{m, q} =\,& \frac{1}{2\pi}\int_{-\frac{L_{\textrm{s}}}{2}}^{\frac{L_{\textrm{s}}}{2}}\int_{-\frac{L_{\textrm{s}}}{2}}^{\frac{L_{\textrm{s}}}{2}}\int_{-\infty}^{\infty}\phi_m(s_x)\frac{A_{\textrm{s}}^2( \kappa_x)}{\gamma(\kappa_x)}e^{-j\kappa_x (s_x-s'_x)}\nonumber\\
&\times \phi^*_q(s'_x)\diff\kappa_x\diff s_x\diff s'_x, \\
\label{eqn: Autocorrelation Matrix Receiver}
 \big[\R_{\textrm{r}}\big]_{n, p}=\,& \frac{1}{2\pi}\int_{-\frac{L_{\textrm{r}}}{2}}^{\frac{L_{\textrm{r}}}{2}}\int_{-\frac{L_{\textrm{r}}}{2}}^{\frac{L_{\textrm{r}}}{2}}\int_{-\infty}^{\infty}\psi_n^*(r_x)\frac{A_{\textrm{r}}^2(k_x)}{\gamma(k_x)}e^{jk_x(r_x-r'_x)}\nonumber\\
 &\times \psi_p(r'_x)  \diff k_x \diff r_x\diff r'_x.
\end{align}
\end{subequations}
Hence, \eqref{eqn: Autocorrelation Matrix elements} can be written in matrix form as
\begin{align}
\label{eqn: Autocorrelation Matrix}
 \R=\R_{\textrm{s}}\otimes \R_{\textrm{r}}\in \mathbb{C}^{n_\textrm{r} n_\textrm{s}\times n_\textrm{r} n_\textrm{s}},
\end{align}
Comparing \eqref{eqn: Autocorrelation Matrix Receiver} with the expression in \cite[Eq.~(71)]{sanguinetti2022wavenumber}, i.e., 
\begin{align}
    \label{eqn: WDM ACM}
    \hspace{-2mm} \big[\R_{\textrm{r}}\big]_{n, p}=\int_{-\frac{L_{\textrm{r}}}{2}}^{\frac{L_{\textrm{r}}}{2}}\int_{-\frac{L_{\textrm{r}}}{2}}^{\frac{L_{\textrm{r}}}{2}}\Gamma_{\textrm{r}}(r_x-r'_x)\psi_n^*(r_x)\psi_p(r'_x) \diff r_x\diff r'_x,
\end{align}
we have the \ac{ACF} at the receiver given by
\begin{align}
\begin{split}
    \Gamma_{\textrm{r}}(r_x) & =\frac{1}{2\pi}\int_{-\infty}^{\infty}\frac{A_{\textrm{r}}^2(k_x)}{\gamma(k_x)}e^{jk_x r_x} \diff k_x \\
    & =\frac{1}{2\pi}\int_{-k}^{k}\frac{A_{\textrm{r}}^2(k_x)}{\sqrt{k^2-k_x^2}}e^{jk_x r_x} \diff k_x,
\end{split}
\label{eqn: autocorrelation1} 
\end{align}
where the last equality follows from \eqref{eqn:gamma} for $k_x\in \setD$. Applying the change of variable $k_x \! = \!k\cos\theta_{\textrm{r}}$, \eqref{eqn: autocorrelation1} simplifies to
\begin{align}
    \label{eqn: autocorrelation2}    
    \Gamma_{\textrm{r}}(r_x)=\frac{1}{2\pi}\int_{0}^{\pi}A_{\textrm{r}}^2(\theta_{\textrm{r}})e^{jk\cos\theta_{\textrm{r}} r_x} \diff \theta_{\textrm{r}}.
\end{align}
The \ac{PSD} at the receiver is obtained as the Fourier transform of the corresponding \ac{ACF} in the wavenumber domain as
\begin{align}
    \label{eqn: Fourier Transform}
    S_{\textrm{r}}(k_x)=\int_{-\infty}^{\infty}\Gamma_{\textrm{r}}(r_x)e^{-jk_x r_x} \diff r_x.
\end{align}
Plugging \eqref{eqn: autocorrelation1} into \eqref{eqn: Fourier Transform} and rearranging the order of integration yields
\begin{align}
    \label{eqn: Fourier Transform1}
    S_{\textrm{r}}(k_x)=\frac{1}{2\pi}\int_{-k}^{k}\frac{A_{\textrm{r}}^2(k'_x)}{\gamma(k'_x)}\diff k'_x\int_{-\infty}^{\infty}e^{j(k'_x-k_x)r_x} \diff r_x.
\end{align}
Now, using the relation $\int_{-\infty}^{\infty}e^{j(k'_x-k_x)r_x} \diff r_x=2\pi\delta(k'_x-k_x)$, \eqref{eqn: Fourier Transform1} simplifies to
\begin{align}
    \label{eqn: Fourier Transform2}
    S_{\textrm{r}}(k_x)=\frac{A_{\textrm{r}}^2(k_x)}{\gamma(k_x)} \mathbbm{1}_{\setD}(k_{x}),
\end{align}
which connects the \ac{PSD} and \ac{PSF}.

Under the assumption of separable scattering, we have $\sigma^2(n, m)=\sigma_{\textrm{s}}^2(m)\sigma_{\textrm{r}}^2(n)$, where $\sigma_{\textrm{s}}^2(m)$ and $\sigma_{\textrm{r}}^2(n)$ are the transmit and receive variances, respectively. Therefore, employing \eqref{eqn: Channel H_nm discrete without sinc}--\eqref{eqn: Autocorrelation}, we express the spatial autocorrelation as
\begin{align}
\label{eqn: Autocorrelation angular1 }
    R_{mn, qp}= L_{\textrm{s}} L_{\textrm{r}} \sigma_{\textrm{r}}^2(n)\sigma_{\textrm{s}}^2(m)\delta[n-p]\delta[m-q],
\end{align}
which can be written in matrix form as
\begin{align}
\label{eqn: Autocorrelation Matrix different}
 \R= \diag({\sigmab}_{\textrm{s}} \odot {\sigmab}_{\textrm{s}})\otimes \diag({\sigmab}_{\textrm{r}} \odot {\sigmab}_{\textrm{r}}),
\end{align}
where ${\sigmab}_{\textrm{s}}\in \Real^{n_{\textrm{s}}}_+$ and ${\sigmab}_{\textrm{r}}\in \Real^{n_{\textrm{r}}}_+$ collect the scaled standard deviations $\left\{\sqrt{L_{\textrm{s}}}\sigma_{\textrm{s}}(m): m\in \setE_{\textrm{s}}\right\}$ and $\left\{\sqrt{L_{\textrm{r}}}\sigma_{\textrm{r}}(n): n\in \setE_{\textrm{r}}\right\}$, respectively. Finally, comparing \eqref{eqn: Autocorrelation Matrix} with \eqref{eqn: Autocorrelation Matrix different}, we have
\begin{subequations}
\label{eqn: Autocorrelation Matrix Source Diff and Receiver Diff}
\begin{align}
\R_{\textrm{s}}&=\diag({\sigmab}_{\textrm{s}} \odot {\sigmab}_{\textrm{s}}),\\
\R_{\textrm{r}}&=\diag({\sigmab}_{\textrm{r}} \odot {\sigmab}_{\textrm{r}}).
\end{align}
\end{subequations}
With \eqref{eqn: Autocorrelation Matrix Source Diff and Receiver Diff}, the \ac{WDM}-based \ac{NLoS} channel matrix can be obtained as
\begin{align}
    \label{eqn: Complete channel matrix}
    {\H}={\R}_{\textrm{r}}^{\frac{1}{2}} \W {\R}_{\textrm{s}}^{\frac{1}{2}} \in \mathbb{C}^{n_{\textrm{r}} \times n_{\textrm{s}}},
\end{align}
where $\W\in \mathbb{C}^{n_{\textrm{r}} \times n_{\textrm{s}}}$ is a random matrix with \ac{i.i.d.} $\setN_{\mathbb{C}}(0,1)$ entries.

\subsection{Characterization of the Angular-Domain NLoS Channel}
\label{sec: Calculation of Channel Coefficients for WDM and holographic channel}

Characterizing $H_{\textrm{a}}(n, m)$ is essential for deriving expressions for the \ac{WDM}-based \ac{NLoS} channel matrix. Since we have $H_{\textrm{a}}(n, m) \sim \setN_{\mathbb{C}}(0,\sigma^2(n, m))$ from \eqref{eqn: Discrete Angular}, the following analysis focuses on characterizing $\sigma^2(n, m)$.
Equation \eqref{eqn: Discrete NLoS} represents the Fourier series expansion of $h(\r,\s)$ under the assumption that the normalized lengths at the source and receiver are large, i.e., $\frac{L_{\textrm{s}}}{\lambda}\gg1$ and $\frac{L_{\textrm{r}}}{\lambda}\gg1$. In this setting, $\sigma^2(n, m)$ is obtained as the sampled value of the \ac{PSD} at the point $\big(\frac{2\pi}{L_{\textrm{r}}}n, \frac{2\pi}{L_{\textrm{s}}}m\big)$ in the wavenumber domain \cite[Eq.~(177)]{van2004detection}. Since separable scattering implies $\sigma^2(n, m)=\sigma_{\textrm{s}}^2(m)\sigma_{\textrm{r}}^2(n)$ (see Section~\ref{sec: Autocorrelation of Channel H_{nm}}), we focus on modeling the variance at the receiver, as the procedure for the source follows analogously.

The variance at the receiver after performing a change of variable similar to \eqref{eqn: autocorrelation2} and utilizing \eqref{eqn: Fourier Transform2} is given by
\begin{align}
    \label{eqn: sigma at receiver change of variable} \sigma_{\textrm{r}}^2(n) = \frac{1}{2\pi}\int_{{\setT}_{\textrm{r}}(n)}A^2_{\textrm{r}}(\theta_{\textrm{r}})\diff\theta_{\textrm{r}},
\end{align}
with $\setT_{\textrm{r}}(n) = \big\{\arccos\big({\frac{\lambda}{L_{\textrm{r}}}(n+1)}\big), \arccos\big({\frac{\lambda}{L_{\textrm{r}}}n}\big)\big\}$; the detailed steps are presented in \cite{praj25journal}. The \ac{PSF} is determined by $\tilde{A}_{\textrm{r}}^2(\theta_{\textrm{r}})=\frac{{A}_{\textrm{r}}^2(\theta_{\textrm{r}})}{2\pi}$, which is modeled as a mixture of \ac{2D} \ac{vMF} distributions \cite{pizzo2022spatial},~i.e.,
\begin{align}
\label{eqn:PowerSpecralFactorLine} \tilde{A}_{\textrm{r}}^2(\theta_{\textrm{r}})=\sum_{\ell=1}^{C} w_{\ell} p_{\ell} (\theta_{\textrm{r}}),
\end{align}
where $C$ denotes the number of scattering clusters, $p_{\ell} (\theta_{\textrm{r}})$ is the \ac{2D} \ac{vMF} distribution for the $\ell$-th cluster, and the positive weights are such that $\sum_{\ell} w_{\ell}=1$. Let $I_0(\cdot)$ and $I_1(\cdot)$ denote the modified Bessel functions of the first kind and order zero and one, respectively. The \ac{2D} \ac{vMF} distribution is defined as~\cite{mardia2000directional}

$ $ \vspace{-10mm}

\begin{align}
    \label{eqn:2DdvMF}
   \hspace{-2.5mm} p_{\ell}(\theta_{\textrm{r}})\! = \! \frac{1}{2\pi I_0(\alpha_{\ell})}\exp\big(\alpha_{\ell}\cos{(\theta_{\textrm{r}} \! - \! \bar{\theta}_{\textrm{r},\ell}})\big), \  \theta_{\textrm{r}}\in[-\pi, \pi),
\end{align}
where $\alpha_{\ell}\in \Real^{+}$ is the concentration parameter and $\bar{\theta}_{\textrm{r},\ell}\in[-\pi, \pi)$ is the mean angle of the $\ell$-th cluster. For a given normalized circular variance $\nu_{\ell}^2\in[0, 1]$, we compute $\alpha_{\ell}$ using the fixed-point equation $\nu_{\ell}^2=1-\big(\frac{I_1(\alpha_{\ell})}{I_{0}(\alpha_{\ell})}\big)^2$.
 
We now consider the case of isotropic scattering, obtained by setting $C=1$ with $\alpha_{1}=0$, which implies $\nu_{1}^2=1$. Considering only the forward-traveling wave, i.e., $\theta_{\textrm{r}}\in[0, \pi)$, and using \eqref{eqn:PowerSpecralFactorLine}--\eqref{eqn:2DdvMF}, we have
\begin{align}
    \label{eqn: one isotropic}
    \tilde{A}_{\textrm{r}}^2(\theta_{\textrm{r}})=\frac{1}{\pi}, \  \theta_{\textrm{r}}\in[0, \pi).
\end{align}
In this case, since \eqref{eqn: one isotropic} no longer depends on $\bar{\theta}_{\textrm{r}}$, \eqref{eqn: autocorrelation2} can be expressed as $\Gamma_{\textrm{r}}(r_x)={J_0(k r_x)}$, where $J_0(\cdot)$ is the Bessel function of the first kind and order zero, and the corresponding \ac{PSD} is given by $S_{\textrm{r}}(k_x)=\frac{2}{\sqrt{k^2-k_x^2}}$, $|k_x|\leq k.$
Hence, under isotropic scattering, the \ac{PSD} and \ac{ACF} coincide with the classical Jakes' isotropic model.

\section{Numerical Results}
\label{sec: Numerical Results}

\begin{figure}[t!]
    \centering
    \input{ASILOMAR_Conference/ASILOMAR_Results/files_tex/PSF_NonIso_Asilomar}
        \caption{Illustration of $\tilde{A}_\mathrm{r}^2(\theta_\mathrm{r})$.}
        \label{fig:Fig3a} \vspace{-1mm}
\end{figure}
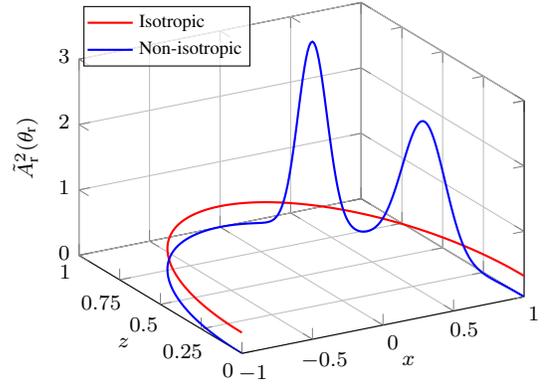

We consider holographic lines with lengths $L_{\textrm{s}} = L_{\textrm{r}} =128\lambda$ and wavelength $\lambda=0.01$~m (corresponding to a carrier frequency of $30$~GHz). We compare the \ac{WDM}-based \ac{NLoS} channel with \ac{i.i.d.} Rayleigh fading and Jakes’ isotropic models, which are obtained by spatially sampling the holographic lines according to the Nyquist criterion, i.e., with $\frac{\lambda}{2}$ spacing. To simplify the analysis, we assume symmetric scattering between the source and receiver, i.e., $\tilde{A}_\textrm{s}^2(\theta_\textrm{s})=\tilde{A}_\textrm{r}^2(\theta_\textrm{r})$, which yields $\sigma_\textrm{r}^2(n)=\sigma_\textrm{s}^2(m)$. For non-isotropic scattering, we consider $C=2$ scattering clusters with mean angles $\bar{\theta}_{\textrm{r},1}=30^\circ$ and  $\bar{\theta}_{\textrm{r},2}=60^\circ$, with normalized circular variances $\nu_{1}^2=0.01$ and $\nu_{2}^2=0.005$, respectively, and weights $w_{1}=w_{2}=\frac{1}{2}$. Fig.~\ref{fig:Fig3a} illustrates $\tilde{A}_\textrm{r}^2(\theta_\textrm{r})$ for isotropic and non-isotropic scattering obtained using \eqref{eqn: one isotropic} and \eqref{eqn:PowerSpecralFactorLine}--\eqref{eqn:2DdvMF}, respectively, for the forward-traveling waves, i.e., $\theta_{\textrm{r}}\in[0, \pi)$.

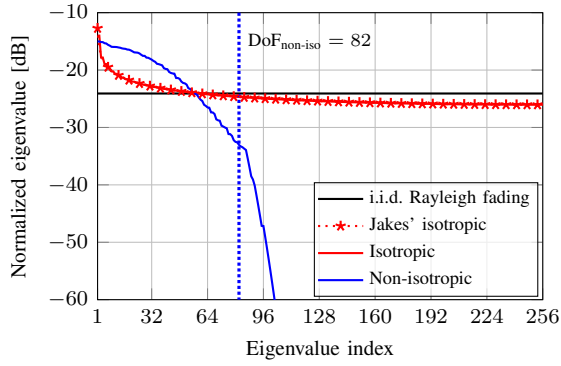
\begin{figure}[t!]
    \centering
    \input{ASILOMAR_Conference/ASILOMAR_Results/files_tex/Rr_eigen_0.5lambda}
    \caption{Normalized eigenvalues of $\mathbf{R}_{\mathrm{r}}$.}
    \label{fig:Fig6}
\end{figure}

\begin{figure}[t!]
    \centering
    \input{ASILOMAR_Conference/ASILOMAR_Results/files_tex/SNR_SE_NLoS_CSIT}
    \caption{Ergodic capacity versus total transmit power.}
    \label{fig:Fig9} \vspace{-1mm}
\end{figure}
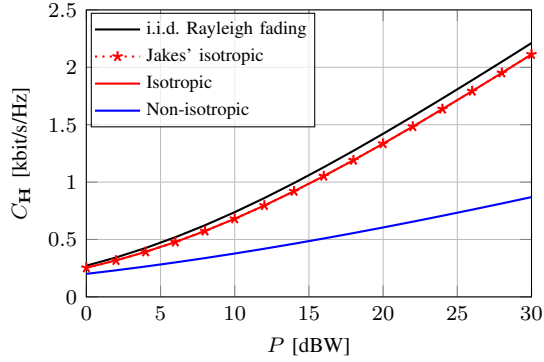

We examine the \ac{DoF} defined as follows: for isotropic scattering, they are the minimum number of non-zero coupling coefficients in $\setE_{\textrm{s}}$ and $\setE_{\textrm{r}}$ required to represent $h(\r,\s)$ over the linear regions $\setL_{\textrm{s}}$ and $\setL_{\textrm{r}}$ \cite{pizzo2020degrees}, i.e., $\textrm{DoF}_\textrm{iso}=\mathrm{min}(n_{\textrm{s}}, n_{\textrm{r}})$; for non-isotropic scattering, they coincide with the \ac{DoF} of the underlying random process \cite{franceschetti2017wave}, i.e.,

$ $ \vspace{-10mm}

\begin{align}
    \label{eqn: DoF Non isotropic def}
    \hspace{-2mm}\textrm{DoF}_\textrm{non-iso} \! = \! \min \biggl\{ n_{\textrm{s}}' : \hspace{-0.7mm}\sum_{i=1}^{n_{\textrm{s}}'} \hspace{-0.7mm}\sigma_{\textrm{s},i}^{2} \! \geq \! 1 \! - \! \epsilon, \, n_{\textrm{r}}' : \hspace{-0.9mm}\sum_{i=1}^{n_{\textrm{r}}'} \hspace{-0.9mm}\sigma_{\textrm{r},i}^{2} \! \geq \! 1 \! - \! \epsilon \biggr\},
\end{align}
where $\epsilon$ specifies the desired level of accuracy and $\{\sigma_{\textrm{s},i}^{2}\}_{i=1}^{n_{\textrm{s}}}$ and $\{\sigma_{\textrm{r},i}^{2}\}_{i=1}^{n_{\textrm{r}}}$ represent $\big\{\sigma_{\textrm{s}}^{2}(m): m \in \setE_{\textrm{s}}\big\}$ and $\big\{\sigma_{\textrm{r}}^{2}(n): n \in \setE_{\textrm{r}}\big\}$, respectively, obtained as in \eqref{eqn: sigma at receiver change of variable} and sorted in decreasing order. As done in \cite{pizzo2022fourier}, we set $\epsilon = 0.3\%$ according to the three-sigma rule of the Gaussian distribution, stating that about $99.7\%$ of the values lie within three standard deviations. Fig.~\ref{fig:Fig6} plots the eigenvalues of the receive spatial autocorrelation matrix $\R_{\textrm{r}}$ (normalized by its trace). For \ac{i.i.d.} Rayleigh fading, the channel entries are mutually independent, leading to identical eigenvalues of $\R_{\textrm{r}}$. The closest physically meaningful counterpart is Jakes' isotropic model, whose spatial autocorrelation matrix is obtained from the spatial sampling of $J_0(k r_x)$. The normalized eigenvalues under isotropic scattering coincide with those of Jakes' isotropic model, whereas the non-isotropic case exhibits a significantly steeper decay, indicating stronger spatial correlation. In terms of \ac{DoF}, we have $\textrm{DoF}_{\textrm{iso}}=256$ and $\textrm{DoF}_{\textrm{non-iso}}=82$.

Considering \eqref{eqn: input-output relationship} and assuming perfect channel state information at both the source and receiver, the ergodic capacity (measured in bits/s/Hz) is given by
\begin{align}
    \label{eqn: Spectral Efficiency CSIT}  
    C_{\H}
    = \Exp \bigg[ \sum_{i=1}^{N}
      \log_{2} \left(1+\frac{P_{i}}{\chi^{2}}
      \varrho_{i}(\H\H^{\herm})\right) \bigg],
\end{align}
where $\varrho_{i}(\cdot)$ denotes the $i$-th eigenvalue of $\H\H^{\herm}$ sorted in decreasing order, $P_i$ is the transmit power allocated to the $i$-th eigenmode, computed via water-filling for each channel realization, and $P = \sum_{i=1}^{N} P_i$ represents the total transmit power. For simplicity, we assume $\chi^2=0$~dBW and compute \eqref{eqn: Spectral Efficiency CSIT} by averaging over $500$ independent channel realizations. Fig.~\ref{fig:Fig9} depicts the ergodic capacity against the total transmit power. With water-filling power allocation, the capacity reflects the eigenvalue distribution of $\R_{\textrm{r}}$. The curve corresponding to Jakes' isotropic model closely matches that of isotropic scattering. The non-isotropic case yields the lowest capacity due to the stronger spatial correlation and the correspondingly smaller number of dominant eigenmodes. The capacity under \ac{i.i.d.} Rayleigh fading also approaches that of the isotropic and Jakes' models.

\section{Conclusions}
\label{sec: Conclusions}

We analyzed the \ac{WDM}-based \ac{NLoS} channel in holographic \ac{MIMO} by characterizing its spatial autocorrelation. In particular, we showed that applying \ac{WDM} to the \ac{NLoS} channel leads to its angular-domain representation. Furthermore, particularizing the analysis to isotropic scattering recovers Jakes’ isotropic model. The results highlight that \ac{WDM} serves not only as a multiplexing technique but also as an effective analytical tool for studying holographic \ac{MIMO} channels. Future work may explore extensions to the multi-user setting.

\addcontentsline{toc}{chapter}{References}
\bibliographystyle{IEEEtran}
\bibliography{refs_abbr, refs}

\end{document}

%% file: Definitions.tex
\newcommand{\e}{\mathbf{e}}

\renewcommand{\j}{\mathbf{j}}
\renewcommand{\k}{\mathbf{k}}

\renewcommand{\r}{\mathbf{r}}
\newcommand{\s}{\mathbf{s}}

\newcommand{\x}{\mathbf{x}}

\renewcommand{\H}{\mathbf{H}}

\newcommand{\R}{\mathbf{R}}

\newcommand{\W}{\mathbf{W}}

\newcommand{\kappab}{\boldsymbol{\kappa}}

\newcommand{\sigmab}{\boldsymbol{\sigma}}

\newcommand{\psib}{\boldsymbol{\psi}}

\newcommand{\setD}{\mathcal{D}}
\newcommand{\setE}{\mathcal{E}}

\newcommand{\setL}{\mathcal{L}}

\newcommand{\setN}{\mathcal{N}}

\newcommand{\setT}{\mathcal{T}}

\newcommand{\Compl}{\mbox{$\mathbb{C}$}}
\newcommand{\Real}{\mbox{$\mathbb{R}$}}

\newcommand{\diag}{\mathrm{diag}}

\newcommand{\diff}{\mathrm{d}}
\newcommand{\Exp}{\mathbb{E}}
\newcommand{\herm}{\mathrm{H}}

\newcommand{\tran}{\mathrm{T}}

\newcommand{\card}{\mathrm{card}}

\definecolor{oulu_blue}{HTML}{23408F}
\definecolor{oulu_green}{HTML}{39B54A}

\definecolor{red}{rgb}{1,0,0}
\definecolor{red_magenta}{rgb}{1,0,0.5}
\definecolor{magenta}{rgb}{1,0,1}
\definecolor{blue_magenta}{rgb}{0.5,0,1}
\definecolor{blue}{rgb}{0,0,1}
\definecolor{blue_cyan}{rgb}{0,0.5,1}
\definecolor{cyan}{rgb}{0,1,1}
\definecolor{green_cyan}{rgb}{0,1,0.5}
\definecolor{green}{rgb}{0,1,0}
\definecolor{green_yellow}{rgb}{0.5,1,0}
\definecolor{yellow}{rgb}{1,1,0}
\definecolor{red_yellow}{rgb}{1,0.5,0}

%% file: Acronyms.tex
\begin{acronym}
\acro{1D}{one-dimensional}
\acro{2D}{two-dimensional}
\acro{3D}{three-dimensional}
\acro{ACF}{autocorrelation function}
\acro{AWGN}{additive white Gaussian noise}
\acro{DoF}{degrees of freedom}
\acro{EM}{electromagnetic}
\acro{i.i.d.}{independent and identically distributed}
\acro{ITU}{International Telecommunication Union}
\acro{LoS}{line-of-sight}
\acro{MIMO}{multiple-input multiple-output}
\acro{NLoS}{non-line-of-sight}
\acro{OFDM}{orthogonal frequency-division multiplexing}
\acro{PSD}{power spectral density }
\acro{PSF}{power spectral factor}
\acro{SNR}{signal-to-noise ratio}
\acro{vMF}{von Mises-Fisher}
\acro{WDM}{wavenumber-division multiplexing}
\end{acronym}

%% file: ASILOMAR_Conference/ASILOMAR_Results/files_tex/WDM_ASILOMAR.tex
\begin{tikzpicture}[>=latex]

\small

\def\dh{0.75cm}
\def\dv{0.75cm}
\node[] (center) at (0,0) {};
\node[left=\dh of center] (center_left) {};
\node[above=1.5*\dv of center_left] (top_left) {};
\node[below=1.5*\dv of center_left] (bottom_left) {};
\node[right=\dh of center] (center_right) {};
\node[above=1.5*\dv of center_right] (top_right) {};
\node[below=1.5*\dv of center_right] (bottom_right) {};

\draw[very thick] (top_left) -- (bottom_left) node[midway, right=1pt] {$\j(\s)$};
\draw[very thick] (top_right) -- (bottom_right) node[midway, left=1pt] {$\e(\r)$};

\node[draw, minimum width=0.8cm, minimum height=0.5cm, fill=gray!20, ellipse, rotate=60, transform shape] () at ($(center)+(0,\dv)$) {};
\node[draw, minimum width=0.5cm, minimum height=0.4cm, fill=gray!20, ellipse, rotate=120, transform shape] () at ($(center)+(0,-\dv)$) {};

\node[draw, circle, left=\dh/4 of center_left, align=center, inner sep=0pt] (+_left) {+};
\draw[->] (+_left.east) -- (center_left.center);
\node[right=\dh/4 of center_right, align=center, inner sep=0pt] (cross_right) {};
\draw[-] (center_right.center) -- (cross_right.center);

\node[draw, rectangle, above left=1.25*\dv and \dh/3 of +_left.center, align=center, text width=1cm] (phi_1) {$\phi_1(s_x)$};
\node[draw, rectangle, below left=1.25*\dv and \dh/3 of +_left.center, align=center, text width=1cm] (phi_2) {$\phi_N(s_x)$};
\node[align=center] () at ($(+_left)+(-\dv-0.1cm,0cm)$) {$\vdots$};
\draw[->] (phi_1.east) -- ($(phi_1.east)+(\dh/3,0)$) -- (+_left.north);
\draw[->] (phi_2.east) -- ($(phi_2.east)+(\dh/3,0)$) -- (+_left.south);

\node[draw, rectangle, above right=1.25*\dv and \dh/3 of cross_right.center, align=center, text width=2cm] (psi_1) {$\int(\cdot)\psi_{1}^{*}(r_x)\diff r_x$};
\node[draw, rectangle, below right=1.25*\dv and \dh/3 of cross_right.center, align=center, text width=2cm] (psi_2) {$\int (\cdot)\psi_{N}^{*}(r_x)\diff r_x$};
\node[align=center] () at ($(cross_right)+(\dv+0.6cm,0cm)$) {$\vdots$};
\draw[->] (cross_right.center) -- ($(psi_1.west)+(-\dh/3,0)$) -- (psi_1.west);
\draw[->] (cross_right.center) -- ($(psi_2.west)+(-\dh/3,0)$) -- (psi_2.west);

\node[left=\dv/2 of phi_1, align=center, inner sep=0pt] (x_1) {$x_1$};
\node[left=\dv/2 of phi_2, align=center, inner sep=0pt] (x_2) {$x_N$};
\draw[->] (x_1.east) -- (phi_1.west);
\draw[->] (x_2.east) -- (phi_2.west);

\node[draw, circle, right=\dv/2 of psi_1, align=center, inner sep=0pt] (+_right_1) {+};
\node[draw, circle, right=\dv/2 of psi_2, align=center, inner sep=0pt] (+_right_2) {+};
\node[right=\dv/2 of +_right_1, align=center, inner sep=0pt] (y_1) {$y_1$};
\node[right=\dv/2 of +_right_2, align=center, inner sep=0pt] (y_2) {$y_N$};
\node[below=\dv/2 of +_right_1, align=center, inner sep=0pt] (z_1) {$z_1$};
\node[above=\dv/2 of +_right_2, align=center, inner sep=0pt] (z_2) {$z_N$};
\draw[->] (psi_1.east) -- (+_right_1.west);
\draw[->] (psi_2.east) -- (+_right_2.west);
\draw[->] (+_right_1.east) -- (y_1.west);
\draw[->] (+_right_2.east) -- (y_2.west);
\draw[->] (z_1.north) -- (+_right_1.south);
\draw[->] (z_2.south) -- (+_right_2.north);


\end{tikzpicture}

%% file: ASILOMAR_Conference/ASILOMAR_Results/files_tex/PSF_NonIso_Asilomar.tex
\begin{tikzpicture}
\begin{axis}[
	width=7.5cm, height=6.25cm,
	xmin=-1, xmax=1,
	ymin=0, ymax=1,
	zmin=0, zmax=3,
	view={-30}{30},
	xlabel={$x$},
	ylabel={$z$},
	zlabel={$\tilde{A}_{\textrm{r}}^2(\theta_{\textrm{r}})$},
	x label style={at={(axis description cs:0.74,0.02)}, anchor=north},
	y label style={at={(axis description cs:0.1,0.07)}, anchor=north},
	label style={font=\footnotesize},
	xtick={-1,-0.5,...,1},
	ytick={0,0.25,...,1},
	ztick={0,1,...,3},
	ticklabel style={font=\footnotesize},
	legend style={at={(0.01,0.99)}, anchor=north west, font=\scriptsize, inner sep=1pt, fill opacity=0.75, draw opacity=1, text opacity=1},
	legend cell align=left,
	title style={font=\scriptsize, yshift=-2mm},
	grid=major
]

\addplot3[thick, red] 
table [x=Var1, y=Var2, z=Var4, col sep=comma]{ASILOMAR_Conference/ASILOMAR_Results/files_txt/PAS_Iso.txt};
\addlegendentry{Isotropic}

\addplot3[thick, blue] 
table[x=Var1, y=Var2, z=Var4, col sep=comma]{ASILOMAR_Conference/ASILOMAR_Results/files_txt/PAS_NoNIso.txt};
\addlegendentry{Non-isotropic}

\end{axis}

\end{tikzpicture}

%% file: ASILOMAR_Conference/ASILOMAR_Results/files_tex/Rr_eigen_0.5lambda.tex
\begin{tikzpicture}
   \begin{axis}[
	width=7.5cm,
	height=5.4cm,
	xmin=1, xmax=256,
	ymin=-60, ymax=-10,
	xlabel={Eigenvalue index},
	ylabel={Normalized eigenvalue [dB]},
       xtick={1,32,64,...,256},
      ytick={-60,-50,...,-10},
	xlabel near ticks,
	ylabel near ticks,
	x label style={font=\footnotesize},
	y label style={font=\footnotesize},
	ticklabel style={font=\footnotesize},
	legend style={at={(0.99,0.01)}, anchor=south east},
	legend style={font=\scriptsize, inner sep=1pt, fill opacity=0.75, draw opacity=1, text opacity=1},
	legend cell align=left,
	grid=both,
	title style={font=\scriptsize},
]  
    \addplot[thick, black] 
    table [x=Var1, y=iid, col sep=comma] {ASILOMAR_Conference/ASILOMAR_Results/files_txt/Rr_Eigen_0.5lambda_normalized.txt};
    \addlegendentry{i.i.d. Rayleigh fading}
    \addplot[thick, red, dotted, mark=star, mark options=solid, mark repeat=6]
    table [x=Var1, y=Eigen_values_ClarkedB, col sep=comma] {ASILOMAR_Conference/ASILOMAR_Results/files_txt/Rr_Eigen_0.5lambda_normalized.txt};
    \addlegendentry{Jakes' isotropic}
    \addplot[thick, red]
     table [x=Var1, y=Iso, col sep=comma] {ASILOMAR_Conference/ASILOMAR_Results/files_txt/Rr_Eigen_0.5lambda_normalized.txt};
    \addlegendentry{Isotropic}   
    \addplot[thick, blue]
    table [x=Var1, y=NonIso, col sep=comma] {ASILOMAR_Conference/ASILOMAR_Results/files_txt/Rr_Eigen_0.5lambda_normalized.txt};
    \addlegendentry{Non-isotropic}

\draw[very thick, blue, densely dotted] (axis cs:82,-10) -- (axis cs:82,-60);
\node[black, font=\scriptsize, anchor=west] at (axis cs:82,-15) {$\textrm{DoF}_{\textrm{non-iso}} = 82$};
\end{axis}
\end{tikzpicture}

%% file: ASILOMAR_Conference/ASILOMAR_Results/files_tex/SNR_SE_NLoS_CSIT.tex
\begin{tikzpicture}
   \begin{axis}[
	width=7.5cm,
	height=5.4cm,
	xmin=0, xmax=30,
	ymin=0, ymax=2.5,
    xlabel={$P$ [dBW]},
	ylabel={$C_{\H}$ [kbit/s/Hz]},
    xtick={0,5,10,...,30},
    ytick={0,0.5,1,...,2.5},
	xlabel near ticks,
	ylabel near ticks,
	x label style={font=\footnotesize},
	y label style={font=\footnotesize},
	ticklabel style={font=\footnotesize},
	legend style={at={(0.01,0.99)}, anchor=north west},
	legend style={font=\scriptsize, inner sep=1pt, fill opacity=0.75, draw opacity=1, text opacity=1},
	legend cell align=left,
	grid=both,
	title style={font=\scriptsize},
]  
    \addplot[thick, black] 
    table [x=SNR, y=iid, col sep=comma] {ASILOMAR_Conference/ASILOMAR_Results/files_txt/SNR_SE_NLoS_CSIT_lambda_0.5.txt};
    \addlegendentry{i.i.d. Rayleigh fading}
    \addplot[thick, red, dotted, mark=star, mark options=solid]
    table [x=SNR, y=Jakes, col sep=comma] {ASILOMAR_Conference/ASILOMAR_Results/files_txt/SNR_SE_NLoS_CSIT_lambda_0.5.txt};
    \addlegendentry{Jakes' isotropic}
    \addplot[thick, red]
     table [x=SNR, y=Iso, col sep=comma] {ASILOMAR_Conference/ASILOMAR_Results/files_txt/SNR_SE_NLoS_CSIT_lambda_0.5.txt};
    \addlegendentry{Isotropic}   
    \addplot[thick, blue]
    table [x=SNR, y=NonIso1, col sep=comma] {ASILOMAR_Conference/ASILOMAR_Results/files_txt/SNR_SE_NLoS_CSIT_lambda_0.5.txt};
    \addlegendentry{Non-isotropic}
    \end{axis}
\end{tikzpicture}